\documentclass[usenatbib]{mn2e}
\usepackage{graphicx}
\usepackage{natbib}
\usepackage{amssymb}
\usepackage{aas_macros}
\usepackage{lscape}

\def\degg{\hbox{$\null^\circ$\hskip-3pt.}}

\title[The   tidal   trail  of   NGC~205?]    {The   tidal  trail   of
  NGC~205?\thanks{Based  in part  on observations  made with  the Isaac
  Newton Telescope on the Island of La Palma by the Isaac Newton Group
  in  the Spanish  Observatorio  del  Roque de  los  Muchachos of  the
  Institutode  Astrofisica  de   Canarias.}\thanks{Some  of  the  data
  presented herein were obtained  at the W.M.  Keck Observatory, which
  is  operated  as  a  scientific  partnership  among  the  California
  Institute  of  Technology,  the  University of  California  and  the
  National Aeronautics and  Space Administration.  The Observatory was
  made possible  by the generous  financial support of the  W.M.  Keck
  Foundation.}}     \author   [McConnachie    et    al.]    {A.     W.
  McConnachie${^1}$, M.  J.  Irwin${^1}$,  G.  F.  Lewis${^2}$, R.  A.
  Ibata${^3}$,   S.    C.    Chapman${^4}$,\newauthor   A.    M.    N.
  Ferguson${^5}$, N.  R. Tanvir${^6}$\\ ${^1}$ Institute of Astronomy,
  Madingley  Road,  Cambridge, CB3  0HA,  U.K.\\  ${^2}$ Institute  of
  Astronomy, School  of Physics, A29, University of  Sydney, NSW 2006,
  Australia\\   ${^3}$  Observatoire   de  Strasbourg,   11,   rue  de
  l'Universite,   F-67000,  Strasbourg,  France\\   ${^4}$  California
  Institute  of  Technology,   Pasadena,  CA  91125,  U.S.A.\\  ${^5}$
  Max-Planck-Institut f\"{u}r Astrophysik, Karl-Schwarzschild-Str.  1,
  Postfach 1317, D-85741 Garching, Germany\\ ${^6}$ Physical Sciences,
  Univ. of Hertfordshire, Hatfield, AL10 9AB, U.K.\\}

\begin{document}

\maketitle

\begin{abstract}
  
  Using data  taken as part of  the Isaac Newton  Telescope Wide Field
  Camera  (INT~WFC) survey  of  M31, we  have  identified an  arc-like
  overdensity of  blue, presumably metal-poor, red  giant branch stars
  in  the north-west  quadrant of  M31.  This  feature  is \mbox{$\sim
    1^\circ$}  (15 kpc)  in extent  and  has a  surface brightness  of
  \mbox{$\Sigma_{V'} \simeq  28.5 \pm 0.5$ mags  arcsec$^{-2}$}. The arc
  appears to emanate from the dwarf elliptical galaxy NGC~205, and the
  colour of its red giant branch is significantly different to the M31
  disk  population but  closely resembles  that of  NGC~205.  Further,
  using  data taken  with the  DEep Imaging  Multi-Object Spectrograph
  (DEIMOS) on  Keck II, we  identify the radial velocity  signature of
  this arc.   Its velocity dispersion  is measured to be  \mbox{$\simeq
    10$\,km\,s$^{-1}$},  similar to  that  of the  central regions  of
  NGC~205  and typical  of stellar  streams.  Based  upon  the spatial
  coincidence of  these objects, the surface  brightness, the velocity
  dispersions and the  similarity in colour of the  red giant branches,
  we  postulate  that  the  arc  is  part of  a  stellar  stream,  the
  progenitor of which is NGC~205.

\end{abstract}


\section{Introduction}

Over the  course of the  last decade there  has been a lot  of renewed
interest in  the Local  Group. This  has in part  been due  to several
discoveries     of    nearby     galaxies     by    various     groups
(\citealt{ibata1994,whiting1997,whiting1999,armandroff1998a,armandroff1998b,karachentsev1999})
that  have  been made  possible  because  of  significant advances  in
instrumentation. At the same  time as these observational discoveries,
heirarchical  formation  scenarios  of  structure  formation,  and  in
particular  Cold  Dark  Matter  (CDM),  have  reached  a  sufficiently
detailed level for the Local Group  to begin acting as a laboratory to
test  and constrain  these theoretical  models. Here,  larger galaxies
such as M31 or the Milky Way are postulated to form via the merging of
smaller bodies (\citealt{white1978,searle1978}).

The  Sagittarius  dwarf galaxy,  discovered  by \cite{ibata1994},  was
subsequently  shown to  have a  substantial stellar  stream associated
with                     it                    (eg.                    
\citealt{mateo1998b,majewski1999,dohmpalmer2001,ibata2001b}), a direct
result  of its  interaction with  the gravitational  potential  of the
Milky Way. More recent work has shown that up to 75\% of halo M-giants
actually  belong to  the  Sagittarius dwarf  (\citealt{majewski2003}),
demonstrating the  significant role that this accretion  event has had
on  the make  up of  our Galaxy.   Tidal streams  naturally fit  in to
heirarchical  models of  structure formation,  and evidence  for tidal
disruption has  been found  for many of  the Galactic  dSph satellites
(\citealt{irwin1995}) as  well as for many galaxies  located at higher
redshift (eg. \citealt{chapman2003}).

The Andromeda Galaxy, M31, provides the only example of a giant spiral
galaxy, other  than the  Milky Way, in  which individual stars  can be
easily resolved using  current generations of ground-based technology.
As such, we have conducted a large photometric survey of M31 using the
INT WFC  and the CFH12K  camera on the  Canada-France-Hawaii Telescope
(\citealt{ibata2001a,ferguson2002,mcconnachie2003,irwin2004}  {\it  in
preparation},  hereafter  Papers  I,  II, III  and  IV  respectively),
complete over  a total area of \mbox{40  $\square^\circ$,} ranging out
to  80  kpc  from the  centre  of  this  galaxy  (see the  figures  in
\mbox{Paper~IV).} Study of the structure  of Andromeda is in many ways
easier than for our own Galaxy, as our location external to M31 allows
us to obtain a global view without the projection problems that plague
studies of the structure of the Milky Way.

Initial  results   of  our   survey  were  spectacular,   revealing  a
considerable  amount of unexpected  substructure. Most  significant of
all these was the discovery of  a giant tidal stream, extending out to
some  40\,kpc in  projection  from  the centre  of  M31 (Paper~I).   A
subsequent, deeper  survey using the  CFH12K camera allowed  the three
dimensional position of  the stream to be calculated,  and showed that
this  feature is  over 140\,kpc  in extent  (Paper~III).  A  follow up
kinematic survey using the  DEIMOS instrument on the KECK~II telescope
has  provided radial  velocities for  over 800  stars in  13 different
fields  located at  various  position in  the  halo and  disk of  M31,
concentrating on areas of  known substructure. Initial results of this
survey are detailed in \cite{ibata2004}.

In this {\it Letter}, we  report on the identification of what appears
to be another substantial tidal stream in M31, distinct in both colour
and  kinematics from  the M31  stellar  stream that  has already  been
reported  in \mbox{Papers~I \&  III.} The  projected position  of this
feature,  the  colour  of  its  red  giant  branch  and  its  velocity
dispersion all suggest that it is tidal debris from M31's bright dwarf
elliptical companion, NGC~205.  This  satellite has long been known to
be         tidally        perturbed         by         M31        (eg.
\citealt{hodge1973,kent1987,bender1991,choi2002,demers2003})  and  the
discovery   of  an  associated   stream  will   allow  for   a  better
understanding of its  evolution.  Additionally, if this interpretation
is correct, then  it is yet more evidence of  the ubiquitous nature of
streams  in galactic halos  which, taken  with the  other substructure
present in M31, reveals the complex formation history of this galaxy.

\section{Photometry}

Details of the search strategy, observations and data reduction of our
INT~WFC survey of M31 can be found in Paper~II, and the completed maps
showing  the  global  2D   spatial  distribution  of  various  stellar
populations in M31 can be found in Paper~IV. These stellar populations
were  defined by  isolating different  locii in  the  colour magnitude
diagrams  created   from  our  Johnson  V  (V$^\prime$)   and  Gunn  i
(i$^\prime$) photometry  (Paper~II).  In  Paper~IV, where we  show the
distribution of ``blue'' red giant branch stars, a prominence of stars
is visible  centred at approximately  \mbox{(-0\degg6,1\degg1).} It is
this feature that is analysed here in more depth.

\begin{figure*}
\resizebox{\hsize}{!}{
\includegraphics[width=9.4cm, angle=0]{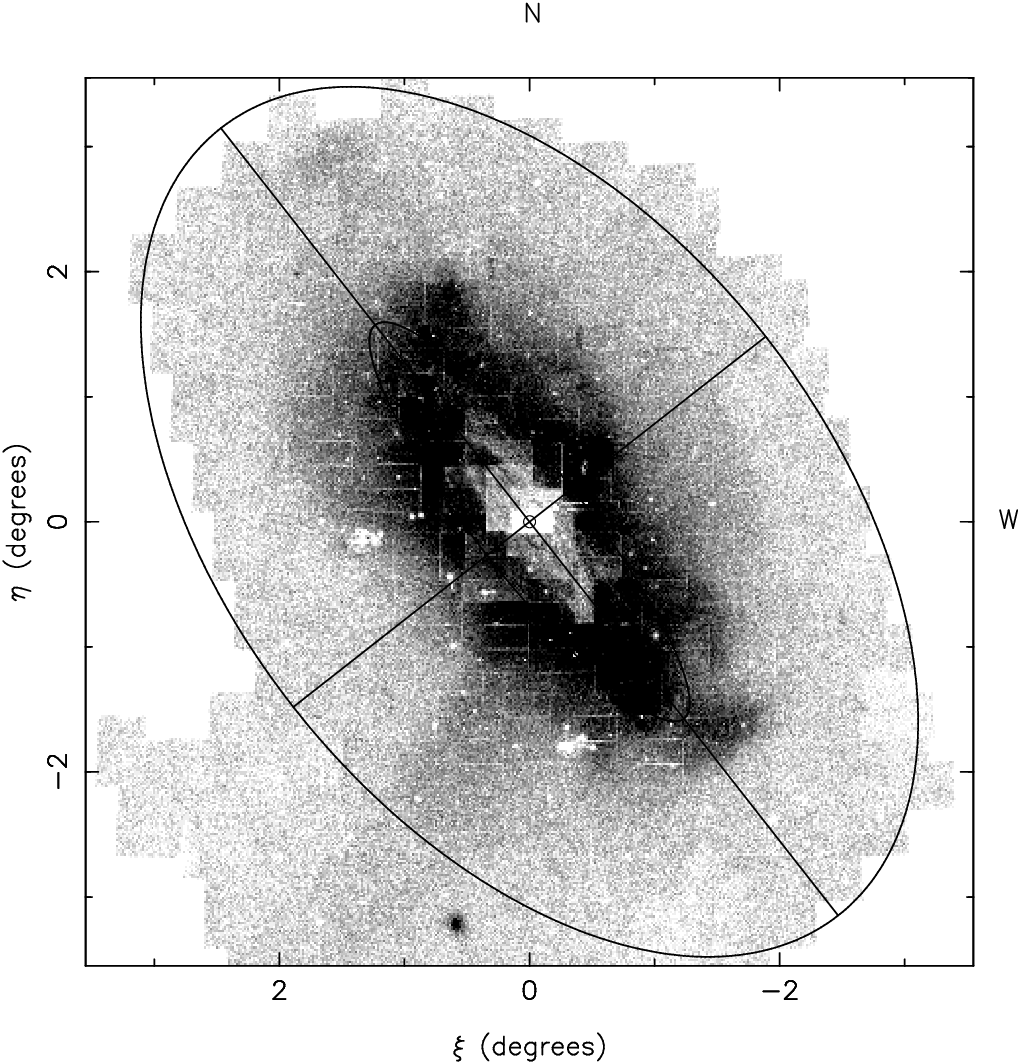}
\hspace*{5mm}
\includegraphics[width=9cm, angle=0]{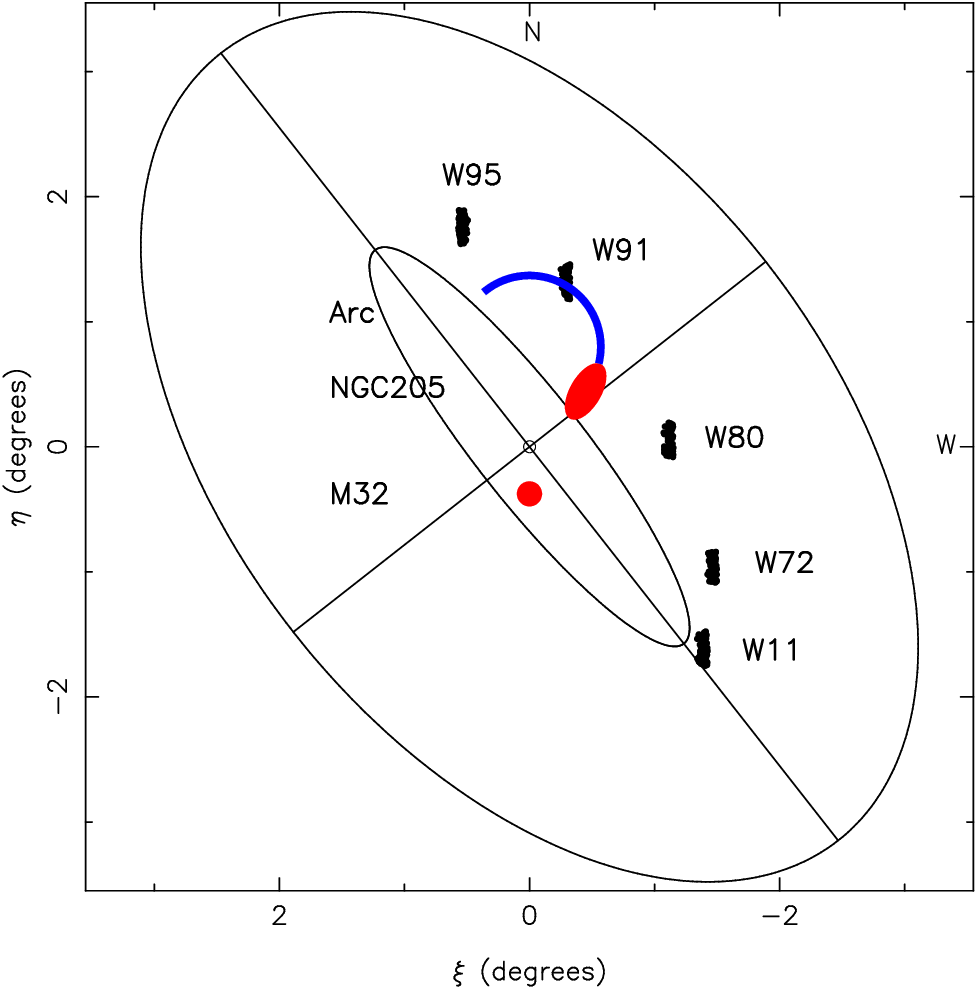}
}
\caption{Left panel: the spatial  distribution of ``bluer'' RGB stars,
defined  by the  the colour  cut described  in the  text. This  cut is
designed  to highlight  the  arc-like density  enhancement centred  at
approximately (-0\degg6,1\degg1). The  dwarf elliptical galaxy NGC~205
is  located at  (-0\degg6,0\degg4).  The  metal poor  dwarf spheroidal
galaxy And~I stands out to the south of the plot, in addition to other
complex substructure  around the outer  SW disk. Right  panel: cartoon
showing the location  of our outer disk DEIMOS  fields with respect to
M32,  NGC~205 and  the stellar  arc. In  both panels,  the $2^{\circ}$
radius ellipse  marks the outer boundary  of the optical  disk of M31,
while the outer ellipse  has a semi-major axis \mbox{$\simeq 55$\,kpc}
and flattening  $0.6$. This corresponds  to the original limit  of the
INT survey.}
\end{figure*}

\begin{figure*}
\begin{center}
\includegraphics[width=7cm, angle=270]{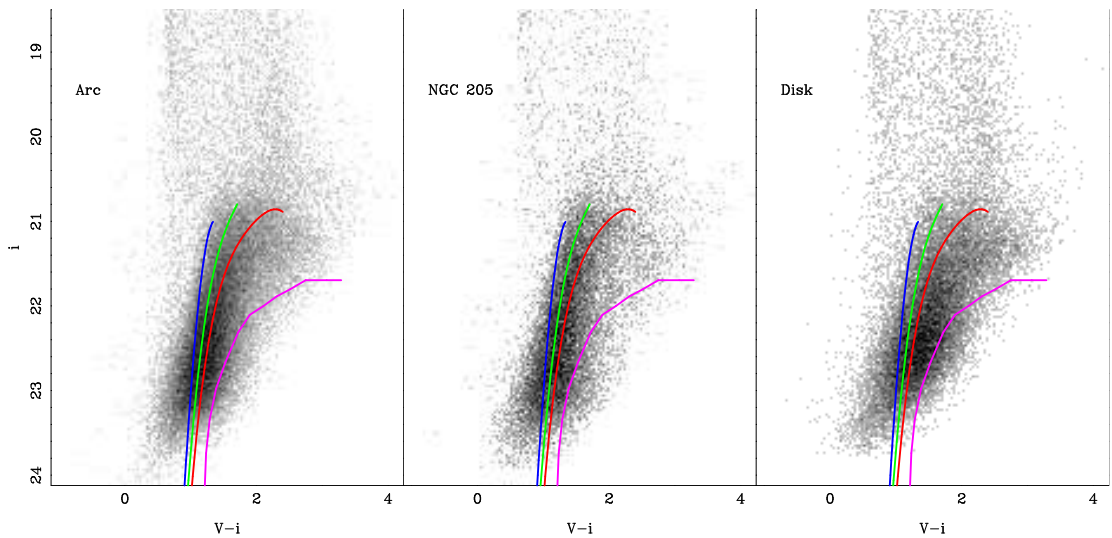}
\caption{Hess  diagrams showing  the CMDs  for the  stellar  arc (left
panel), NGC  205 (middle  panel), and a  typical M31 outer  disk field
(right panel).  Overlaid on these are 4 well studied globular clusters
sequences of  different metallicities. From  left to right,  these are
NGC 6397 (${\rm [Fe/H]} = -1.9$), NGC 1851 (${\rm [Fe/H]} = -1.3$), 47
Tuc  (${\rm [Fe/H]} =  -0.7$) and  NGC 6553  (${\rm [Fe/H]}  = -0.2$).
Both  the arc  and NGC  205 have  an enhanced  bluer red  giant branch
component  compared  to  to  the  M31  field,  located  at  a  similar
galactocentric radius. The enhanced bluer red giant branch in the left
hand panel  appears to be well  represented by a  metallicity of ${\rm
[Fe/H]} \sim  -0.9$, in good agreement  with that derived  for NGC~205
(\citealt{mould1984,mateo1998a}).}
\end{center}
\end{figure*}

The left panel  of Figure 1 shows a map of  ``blue'' RGB stars similar
to  that  shown  in  Paper~IV,  defined  using  a  sample  limited  by
\mbox{20.5   $<$  i$^\prime$  $<$   22.5  mags},   \mbox{24.85  $-   2.85  {\rm \left(V^\prime  -i^\prime\right)} <$  i$^\prime$  $<$ 26.85  $-  2.85 {\rm  \left(V^\prime
      -i^\prime\right)}$.}   An  arc-like  overdensity  is  visible  in  this
figure, with  a projected length of \mbox{$\sim$  1$^\circ$} (15 kpc),
but significantly curved.  It extends to the north of NGC~205 but then
bends back to  the east into the disk of  M31. At these galactocentric
radii we are unable to follow  it further due to the dominant M31 disk
population.  The morphology of  this feature is strongly suggestive of
a stellar stream.  However, the quantity of substructure in our survey
makes it  difficult to catagorise  these features based  on morphology
alone; for  example it could  conceivably also be disrupted  M31 outer
disk.  We estimate  the surface brightness of this  stellar arc in the
V$^\prime$   band  to   be  \mbox{$\Sigma_{{\rm V^\prime}}   \sim  28.5   \pm   0.5$  mags
  arcsec$^{-2}$}.  The Sagittarius stream  has a surface brightness of
\mbox{$\sim  30$ mags  arcsec$^{-2}$}, although  there  is substantial
variation  with azimuthal  angle (\citealt{mateo1998b}),  and  the M31
stream  reported in  Paper~I has  an estimated  surface  brightness of
\mbox{$\sim  30 \pm 0.5$  mags arcsec$^{-2}$.}   The arc  is therefore
typically somewhat brighter than these other remnants, althought still
well within the expected range for stellar streams.

Figure  2 shows a  colour magnitude  diagram (CMD)  for the  arc (left
panel), NGC  205 (middle  panel) and a  typical M31 disk  field (right
panel).   Overlaid  are  4  well  studied  globular  cluster  fiducial
sequences of  different metallicities. From  left to right,  these are
NGC 6397 (${\rm [Fe/H]} = -1.9$), NGC 1851 (${\rm [Fe/H]} = -1.3$), 47
Tuc  (${\rm [Fe/H]}  = -0.7$)  and NGC  6553 (${\rm  [Fe/H]}  = -0.2$)
(\citealt{dacosta1990,sagar1999}).  Both  NGC~205 and the  arc show an
enhanced bluer  red giant branch  component relative to the  M31 field
located at a similar  galactocentric radius. Differential reddening is
unlikely to account for such a  feature. If the stellar arc is somehow
related to the outer  disk of M31, it is not clear  how the colours of
the red giant  braches could be so markedly  different.  Comparison of
the arc's CMD  with the globular cluster tracks shows  that it is well
described   by   a   mean   metallicity   of   [Fe/H]   $\sim   -0.9$.
\cite{mould1984} measured a metallicity for NGC~205, as implied by the
colour of it red giant branch, of [Fe/H] $\sim -0.9 \pm 0.2$ (see also
\citealt{mateo1998a}).  Our INT phototometry therefore suggests a link
between this feature and NGC~205.

\section{Kinematics}

\begin{figure*}
\begin{center}
\includegraphics[width=7cm, angle=270]{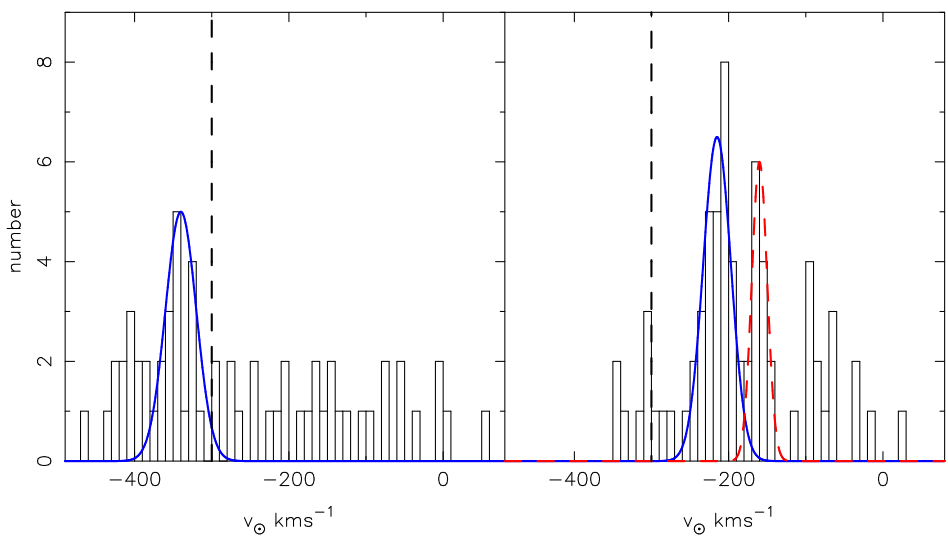}
\caption{The  heliocentric radial velocity  distribution in  W80 (left
  panel) and W91 (right panel). The location of these fields is marked
  on Figure 1.  The vertical dashed line in both panels represents the
  systemic velocity  of M31.  The solid  curves are a  Gaussian fit to
  the   M31   disk   population,    with   a   dispersion   of   $\sim
  20$\,km\,s$^{-1}$. W91 has  an additional component present, centred
  on -160 km\,s$^{-1}$ with a corrected velocity dispersion of $\simeq
  10$\,km\,s$^{-1}$,  which we  attribute to  the stellar  arc (dashed
  curve).}
\end{center}
\end{figure*}

We  have used  the  DEIMOS instrument  on  Keck II  to measure  radial
velocites of  $\simeq$\,80 stars per field  in a total of  13 fields in
various  locations  across  the  extent  of  our  photometric  survey,
selected  to lie  upon various  interesting photometric  features. The
location of the fields in the  outer disk are shown in the right panel
of Figure  1. Details of  the observations and reduction  process, and
early results from this survey, can be found in \cite{ibata2004}.  The
typical uncertainties in the radial velocities that we measure are 5 -
10 kms$^{-1}$. One of our fields (W91)  is centred at an RA and Dec of
\mbox{0h 41m  5,63s,} \mbox{42$^\circ$ 34' 25.1''.}   This samples the
arc  at the  point at  which it  turns to  go back  into the  disk, as
indicated in  Figure 1.  A histogram  of the stars in  this field with
heliocentric radial  velocities in  the range \mbox{$-500  \le v_\odot
\le    80$    km\,s$^{-1}$,}    with    a    Tonry-Davis    coefficent
(\citealt{tonry1979}) greater than 7.5 and an average continuum flux in
the CaII triplet region \mbox{$>$ 50 counts} per pixel (0.31 \AA\ ) is
shown in the right hand panel  of Figure 3.  The left hand panel shows
a comparison  field (W80) located at similar  galactocentric radius on
the  other side  of the  minor axis  (Figure 1).   For  reference, the
systemic     heliocentric     radial     velocity    of     M31     of
$v_\odot\left(\rm{M31}\right)      \simeq      -300$      km\,s$^{-1}$
(\citealt{mateo1998a}) is indicated by the dashed line.

In  both  panels  of  Figure   3,  we  can  identify  various  stellar
components.   The stars  with velocities  $\ge -100$  km\,s$^{-1}$ are
likely foreground  stars in the  halo and disk  of the Milky  Way. The
stars with $v_\odot < -300$  km\,s$^{-1}$ are attributable to the halo
or  bulge  of M31,  and  presumably  have  an equivalent  number  with
$v_\odot > -300$ km\,s$^{-1}$. However,  we attribute the peak at -340
km\,s$^{-1}$ in  W80 and the peak  at -215 km\,s$^{-1}$ in  W91 to the
M31   disk.   As   both   these  fields   are  located   approximately
symmetrically on either side of M31's minor axis, then we would expect
the  disk components  in each  to lie  approximately  symmetrically on
either side  of the M31 systemic  velocity, as we  observe.  The solid
curves  in each  panel  of Figure  3  are Gaussian  fits  to the  disk
component, each with a  velocity dispersion of $\sim 20$\,km\,s$^{-1}$
(uncorrected  for   measuring  errors   of  typically  \mbox{5   -  10
  km\,s$^{-1}$}).  This value is  consistent with our knowledge of the
M31 disk.  In W91 there is an additional velocity component centred on
\mbox{-160  km\,s$^{-1}$}  with  a  corrected  velocity  dispersion  of
\mbox{$\simeq 10$  km\,s$^{-1}$} (dashed curve in Figure  3).  As this
feature   is  unique   to  this   field,  we   attribute  it   to  the
photometrically identified stellar arc.

NGC~205 has  a measured radial  velocity of \mbox{-240  km\,s$^{-1}$.} 
The KECK  radial velocity  field that  we have of  the arc  is located
nearly 1$^\circ$ away  from NGC~205, and shows a  component centred on
\mbox{$-160$  km\,s$^{-1}$} with  a corrected  velocity  dispersion of
\mbox{$\simeq 10$ km\,s$^{-1}$.}  We would expect a substantial velocity
difference between NGC~205  and the arc in this  field whether the two
are related or  not as they are seperated  by \mbox{$\gtrsim 15$ kpc,}
depending  upon the  inclination of  the  arc to  the line  of sight.  
Therefore,  without detailed  modelling or  stellar velocities  in the
region between  our field  and NGC~205, it  is impossible to  draw any
firm  conclusions  regarding their  association  based  upon only  the
magnitude of their radial velocities.  On the other hand, the velocity
dispersion of the arc is  typical of stellar streams and is comparable
to that  which has  been measured for  the central regions  of NGC~205
\mbox{($\sim  14$  km\,s$^{-1}$;}  \citealt{carter1990,peterson1993}),
which  supports a link  between the  two.  We  note that  the velocity
dispersion of  NGC~205 is  known to show  a substantial  increase with
radius, with a velocity dispersion  outside of the nucleus of \mbox{40
  - 50  km\,s$^{-1}$} (\citealt{bender1991}).   This is  indicative of
tidal disruption and so therefore the presence of a tidal stream would
not be surprising.

\section{Discussion}

The quantity of substructure in  M31 makes it difficult to decide upon
the origin  of any single feature (Figure~1;  see also \mbox{Papers~II
\& IV).} Several interpretations as to the origin of this stellar arc
are possible, and here we consider each in turn.

One possibility is  that this feature is a  continuation of the stream
reported in \mbox{Papers~I \&  III.}  This seems unlikely, however, as
the two  structures have significantly different  colours; the stellar
arc is bluer,  suggesting a mean \mbox{[Fe/H] $\sim  -0.9$,} while the
previously  identified  stream is  redder,  with  a mean  \mbox{[Fe/H]
  $\sim$   -0.5}  (Paper~III).    Further,   the  systemic   kinematic
signatures  are  quite   distinct  (\citealt{ibata2004}),  and  so  we
conclude  that  the  stellar   arc  is  unrelated  to  the  previously
identified stellar stream.

A  second  possibility is  that  the  arc  is actually  disrupted  M31
disk. Although this  interpretation is certainly plausible considering
the amount of  disruption evident in M31 we  consider it unlikely, for
two reasons.   Firstly, as Figure  2 demonstrates, this feature  has a
substantially  different  RGB  colour   to  the  outer  disk,  with  a
significantly enhanced  bluer component.  We  would expect the  CMD of
any  disrupted  disk  feature  to  closely resemble  the  CMD  of  the
undisrupted  disk, although  this is  not  the case  here.  While  not
disproving a link between the two, it is not clear how this difference
can be  easily reconciled.   Additionally, the velocity  dispersion of
the  arc ($\simeq  10$ km\,s$^{-1}$)  is  extremely small  for a  disk
component; we measure a M31 disk radial velocity dispersion of $\sim
20$ km\,s$^{-1}$ from Figure 3 and we note that the velocity ellipsoid
in the  Milky Way disk at  the Solar Neighbourhood  is $\sim$ 39:23:20
km\,s$^{-1}$ (\citealt{dehnen1998}).   Again, this suggests  that this
feature is unrelated to the M31 disk.

The  third  possibility  is  that  this  is  a  stellar  stream.   Its
morphological  appearance,  surface  brightness,  and  small  velocity
dispersion are all consistent with this interpretation.  Additionally,
the strong spatial coincidence of the  ``base'' of the arc with NGC~205,
the close similarity  of these objects' red giant  branch colours, and
the  similarity of  the arc's  velocity dispersion  with that  for the
central region of NGC~205 all allude to this object as the progenitor.
As  this  interpretation  is   fully  consistent  with  all  available
photometric and kinematic information,  it is our favoured hypothesis.
To confirm this, radial  velocities in several fields positioned along
the arc are required. If  correct, then a velocity gradient connecting
NGC~205  with the arc in  our field will be  present. This information
will also be  vital, should the connection be  verified, for dynamical
modelling of this intriguing system.

Taking this as our working hypothesis, by counting the excess stars in
the on-stream  regions compared to  off-stream regions, we  estimate a
stellar luminosity  density of  $\sim 1 {\rm  L_\odot}$ arcsec$^{-2}$.
If  we assume  a  value for  the  M/L ratio  in NGC  205  of $\sim  9$
(\citealt{held1990,carter1990,bender1991,peterson1993}), then the mass
density in  this stream is  of order 9 ${\rm  M_\odot}$ arcsec$^{-2}$.
We  estimate the  area  of the  arc  to be  $\sim 0.2  \square^\circ$,
implying  a total  mass  in the  arc  of $\sim  1.8  \times 10^7  {\rm
M_\odot}$, not accounting for projection  effects. The mass of NGC~205
is of  order $7.4  \times 10^8 {\rm  M_\odot}$ (\citealt{mateo1998a}),
and so the visible part of  the stream contains $\sim$ 2.5\,\% of the
mass of NGC~205.

There  has  been  other  evidence  presented in  the  literature  that
suggests NGC~205 is  undergoing tidal disruption. \cite{hodge1973} was
the first to  undertake detailed photometry of this  object and showed
that  its  outer  isophotes  are  twisted,  presumable  due  to  tidal
interactions with  M31. Subsequent  photometry was later  conducted by
\cite{kent1987},    and   he    found   reasonable    agreement   with
\cite{hodge1973}.   Most recently,  \cite{choi2002} have  found strong
evidence for tidal distortion in the isophotes of NGC~205. Figure 9 of
Paper~II is  an isopleth map of  M31 from an  APM scan of a  75 minute
exposure Palomar Schmidt IIIaJ plate  taken by Sydney van den Bergh in
1970,  and it shows  this effect  very clearly.   The increase  in its
velocity   dispersion   as  a   function   of   radius,  reported   by
\cite{bender1991}, also suggests that  there has been some event which
has disturbed the stars in the  outer parts of NGC~205 and caused them
to   have   an  unusually   large   velocity  dispersion.    Recently,
\cite{demers2003} has  shown that the  number of carbon stars  in this
galaxy is low  for its luminosity, and that there  are very few Carbon
stars  located at greater  than $10^{\prime}$  from the  centre, which
suggests that NGC~205 has been tidally stripped by M31.  Our potential
discovery of a  tidal stellar stream is therefore  unsurprising and is
readily  testable   with  further  kinematic  data.    If  the  stream
hypothesis is correct, then the  discovery of this stream will greatly
aid in the  understanding of the evolution of  NGC~205 and provide yet
further  evidence of  the prevalence  of stellar  streams  in galactic
halos and  of the {\it ongoing}  role of accretion  events in galactic
evolution.

\section * {Acknowledgements}
AWM would  like to thank  the University of Sydney  Physics Department
for  their  hospitality  during  a  recent  collaborative  visit.  The
research of AMNF has been supported by a Marie Curie Fellowship of the
European  Community  under  contract number  HPMF-CT-2002-01758.   GFL
acknowledges  support  from  the  Australian  Research  Council  under
Discovery Project DPO343508.

\bibliographystyle{mn2e}
\bibliography{/home/alan/datadisk/latex/papers/references}

\end{document}